\documentclass[sigconf]{acmart}

\copyrightyear{2021}
\acmYear{2021}
\setcopyright{acmcopyright}\acmConference[CIKM '21]{Proceedings of the 30th ACM International Conference on Information and Knowledge Management}{November 1--5, 2021}{Virtual Event, QLD, Australia}
\acmBooktitle{Proceedings of the 30th ACM International Conference on Information and Knowledge Management (CIKM '21), November 1--5, 2021, Virtual Event, QLD, Australia}
\acmPrice{15.00}
\acmDOI{10.1145/3459637.3482392}
\acmISBN{978-1-4503-8446-9/21/11}

\usepackage{amsmath}
\usepackage{amsfonts}
\usepackage{algorithmic}
\usepackage{graphicx}
\usepackage{textcomp}
\usepackage{xcolor}
\usepackage{multirow}
\usepackage[labelformat=simple]{subcaption}
\usepackage{booktabs}
\usepackage{multirow}
\usepackage[ruled,vlined]{algorithm2e}
\usepackage{pifont}
\usepackage{enumitem}

\newcommand{\name}{SBGNN~}
\newcommand{\V}{\mathcal{V}}
\newcommand{\E}{\mathcal{E}}

\newcommand{\eg}{e.g., }
\newcommand{\ie}{i.e., }
\newcommand{\vpara}[1]{\vspace{0.05in}\textbf{#1 }}

\newcommand{\secref}[1]{Section~\ref{#1}}

\newcommand{\figref}[1]{Figure~\ref{#1}}
\newcommand{\tableref}[1]{Table~\ref{#1}} 
 
\newcommand{\algref}[1]{Algorithm~\ref{#1}} 
\newcommand{\lsp}{~\textsc{Link Sign Prediction}~}

\newcommand{\citepp}[1]{~\cite{#1}}

\begin{document}
\fancyhead{}

\title{Signed Bipartite Graph Neural Networks}


\author{Junjie Huang${}^{1,3}$, Huawei Shen${}^{1,3,*}$, Qi Cao${}^{1}$, Shuchang Tao${}^{1,3}$, Xueqi Cheng${}^{2}$}
\affiliation{
  \institution{$^{1}$Data Intelligence System Research Center, }
  \city{Institute of Computing Technology, Chinese Academy of Sciences, Beijing}
  \country{China}
}
\affiliation{
  \institution{ $^{2}$CAS Key Laboratory of Network Data Science and Technology, }
  \city{Institute of Computing Technology, Chinese Academy of Sciences, Beijing}
  \country{China}
}
\affiliation{
  \institution{$^{3}$University of Chinese Academy of Sciences, Beijing, China}
   \country{}
}
\email{{huangjunjie17s, shenhuawei, caoqi, taoshuchang18z, cxq}@ict.ac.cn}

\renewcommand{\shortauthors}{Huang, et al.}

\begin{abstract}

Signed networks are such social networks having both positive and negative links.
A lot of theories and algorithms have been developed to model such networks (\eg balance theory). 
However, previous work mainly focuses on the unipartite signed networks where the nodes have the same type.
Signed bipartite networks are different from classical signed networks, which contain two different node sets and signed links between two node sets.
Signed bipartite networks can be commonly found in many fields including business, politics, and academics, but have been less studied.
In this work, we firstly define the signed relationship of the same set of nodes and provide a new perspective for analyzing signed bipartite networks.
Then we do some comprehensive analysis of balance theory from two perspectives on several real-world datasets.
Specifically, in the peer review dataset, we find that the ratio of balanced isomorphism in signed bipartite networks increased after rebuttal phases.
Guided by these two perspectives, we propose a novel \underline{S}igned \underline{B}ipartite \underline{G}raph \underline{N}eural \underline{N}etworks (SBGNNs) to learn node embeddings for signed bipartite networks. 
SBGNNs follow most GNNs message-passing scheme, but we design new message functions, aggregation functions, and update functions for signed bipartite networks.
We validate the effectiveness of our model on four real-world datasets on \lsp task, which is the main machine learning task for signed networks.
Experimental results show that our SBGNN model achieves significant improvement compared with strong baseline methods, including feature-based methods and network embedding methods. 
\let\thefootnote\relax\footnotetext{*Corresponding Author}
\end{abstract}




\maketitle
{\fontsize{8pt}{8pt} \selectfont
\textbf{ACM Reference Format:}\\
Junjie Huang, Huawei Shen, Qi Cao, Shuchang Tao, Xueqi Cheng. 2021. Signed Bipartite Graph Neural Networks.  In {\it Proceedings of the 30th ACM International Conference on Information and Knowledge Management (CIKM '21), November 1–5, 2021, Virtual Event, QLD, Australia.} ACM, New York, NY, USA, 10 pages. https://doi.org/10.1145/3459637.3482392}

\section{Introduction}

According to a well-known Asian proverb, ``There are a thousand Hamlets in a thousand people’s eyes''.
It means that different people have different opinions.
These differences usually include both positive and negative attitudes.
They are reflected in many fields including business, politics, and academics.
For instance, a research paper submitted to CIKM may receive completely opposite reviews from two reviewers. 
Reviewer 1 gives an ``Accept'' decision, while Reviewer 2 chooses the ``Reject'' option.
Understanding and modeling  these differences is a useful perspective on a range of social computing studies (\eg AI peer review~\cite{heaven2018ai} and congressional vote prediction~\cite{karimi2019multi}).
\begin{figure}
    \centering
    \includegraphics[width=0.9\linewidth]{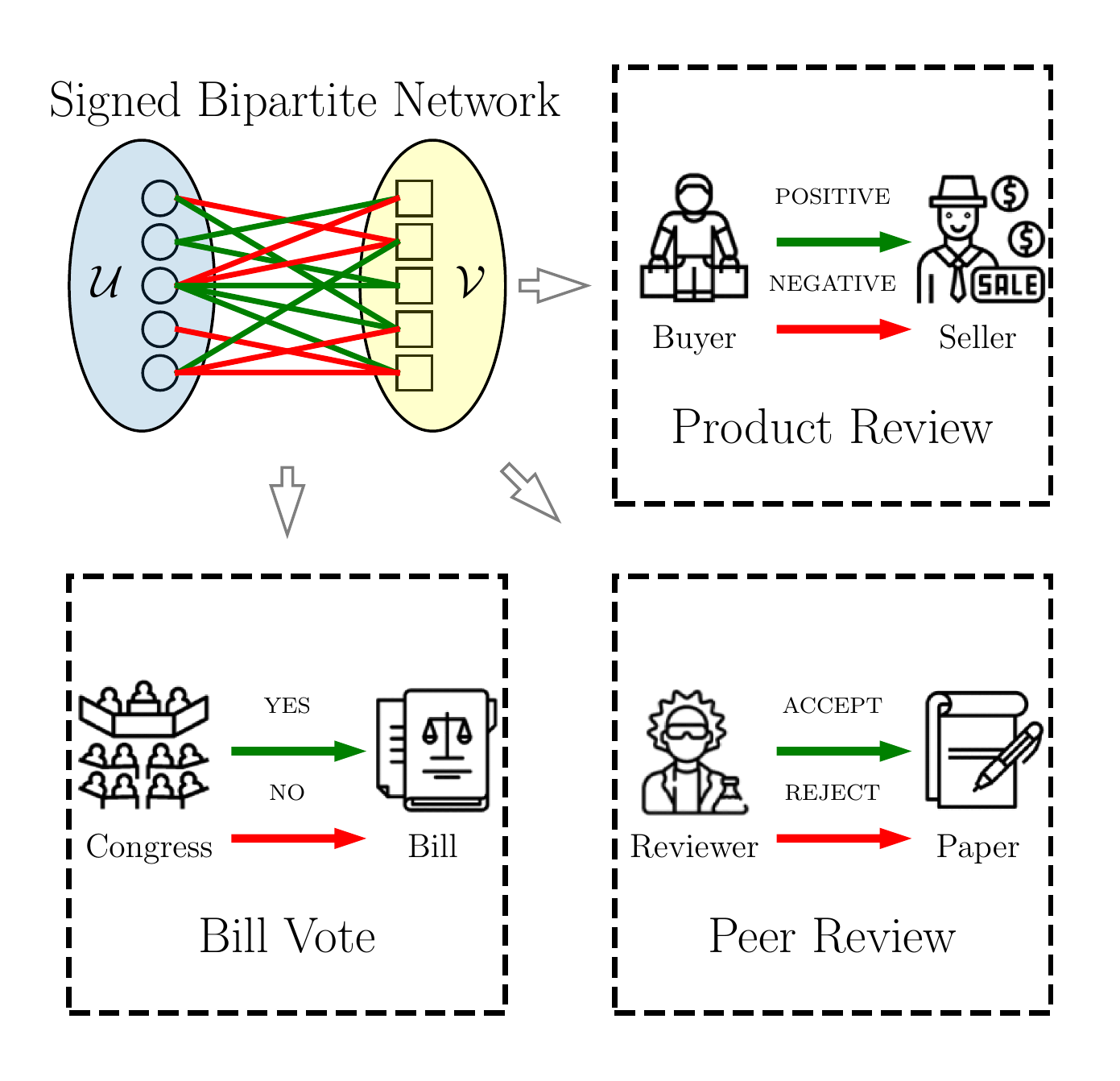}

    \caption{Common application scenarios for signed bipartite networks. }
    \label{fig:application}
    \vspace{-15pt}
\end{figure}
\figref{fig:application} shows some common application scenarios for signed bipartite networks, including product review, bill vote, and peer review.
Some opinions can be viewed as positive relationships, such as favorable reviews on products, supporting the bill, accepting a paper, and so on.
Meanwhile, some opinions are negative links that indicate negative reviews, disapproving a bill, rejecting a paper, and so forth.
These scenarios can be modeled as signed bipartite networks, which include two sets of nodes (\ie $\mathcal{U}$ and $\mathcal{V}$) and the links with positive and negative relationships between two sets.
Compared with unsigned bipartite networks, the links of signed bipartite networks are more complicated, including two opposite relationships (\ie positive and negative links).
Besides, previous works on signed networks only focus on unbipartite signed networks, which are networks that have a single node type~\cite{facchetti2011computing}.
Different node types in signed bipartite networks represent different things.
Modeling signed bipartite networks is a promising and challenging research field.

With modeling the above scenarios into signed bipartite net-works, we do social network analysis on real-world datasets and use advanced graph representation learning methods to model them. 
In signed networks, balance analysis is one of the key research problems in signed graph modeling~\cite{huang2021signlens}.
A common balance analysis method is to count the number of balanced signed triangles in unbipartite signed networks~\cite{leskovec2010signed}.
For signed bipartite networks, \cite{derr2019balance} defines the signed butterfly isomorphism, and uses it to analyze the balance in signed bipartite networks.
But signed butterfly isomorphism may be missing due to data sparsity.
In this paper, we offer a new perspective for analyzing balance theory on signed bipartite networks.
By sign construction, we construct the links between the nodes in the same set and count the signed triangles for the links between the nodes in the same set.
We analyze the balance theory of signed bipartite networks from both two perspectives and explore the balance change in the peer review scenario.
We find that after rebuttal, the balance of the review signed bipartite network increased.
In addition to social network analysis, graph representation learning is another important tool.
Graph Neural Networks (GNNs) have achieved state-of-art results in graph representation learning.
Combing two perspectives, we propose a new \underline{S}igned \underline{B}ipartite \underline{G}raph \underline{N}eural \underline{N}etworks (SBGNNs). 
This model follows the message passing scheme, but we redesign the message function, aggregation function, and update function.
Our SBGNN model achieves the state-of-art results on \lsp, which is the main machine learning task in signed networks~\cite{derr2020link}.
To the best of our knowledge, none of the existing GNN methods has paid special attention to signed bipartite networks. 
It’s the first time to introduce GNNs to signed bipartite networks.
To summarize, the major contributions of this paper are as follows: 
\begin{itemize}[leftmargin=*]
    \item By defining the signed relationship of the same set of nodes (\eg agreement/disagreement), we provide a new perspective for analyzing signed bipartite networks, which can measure the unbalanced structure of the signed bipartite networks from the same set of nodes.
    \item Combining two perspectives, we introduce a new layer-by-layer SBGNN model. 
    Via defining new message functions, aggregation functions, and update functions, SBGNNs aggregate information from neighbors in different node sets and output effective node representations.
    \item We conduct \lsp experiments on four real-world signed bipartite networks including product review, bill vote, and peer review. Experimental results demonstrate the effectiveness of our proposed model.
\end{itemize}

\section{Related Work}
\label{sec:related_work}

\subsection{Signed Graph Modeling}
Signed  networks are such social networks having both positive and negative links\citepp{easley2010networks}.
Balance theory~\cite{heider1944social, cartwright1956structural} is the fundamental theory in the signed network field~\cite{kirkley2019balance}.
For classical signed networks, signed triangles are the most common way to measure the balance of signed networks~\cite{szell2010multirelational}.
Distinct from homogeneous networks, there are two types of nodes in in bipartite networks.
For signed bipartite networks, \cite{derr2019balance} conducts the comprehensive analysis on balance theory in signed bipartite networks, using the smallest cycle in signed bipartite networks (\ie signed butterflies).

To mine signed networks, many algorithms have been developed for lots of tasks, such as community detection\citepp{traag2009community,bonchi2019discovering}, node classification\citepp{tang2016node}, node ranking~\cite{shahriari2014ranking}, and spectral graph analysis\citepp{li2016spectral}.
The \lsp is the main machine learning task for signed networks~\cite{song2015link}.
Modeling balance theory usually leads to better experimental results on \lsp~\cite{huang2021sdgnn}.
For example, \cite{leskovec2010predicting} extracts features by counting signed triangles and achieve good performance in \lsp.
Even with recently signed network embedding methods~\cite{wang2017signed,chen2018bridge,mara2020csne, javari2020rose}, balance theory will be the important guideline for designing models. 
Specifically, SiNE~\cite{wang2017signed} designs an objective function guided by balance theory to learn signed network embeddings and outperforms feature-based methods.
For signed bipartite networks, how to incorporate balance theory is a research-worthy problem.

\subsection{Graph Representation Learning}
Graph Representation Learning (or Network Representation Learning)  is to learn a mapping that embeds nodes, or entire (sub)graphs, as points in a low-dimensional vector space~\cite{hamilton2017representation}.
The nodes in the graph is represented as node embeddings, which reflect the structure of the origin graph.
The common methods for graph representation learning include matrix factorization-based methods~\cite{ou2016asymmetric}, random-walk based algorithms~\cite{perozzi2014deepwalk,tang2015line,grover2016node2vec}, and graph neural networks.
Specifically, Node2vec~\cite{grover2016node2vec} extends DeepWalk~\cite{perozzi2014deepwalk} by performing biased random walks to generate the corpus of node sequences; and it efficiently explores more diverse neighborhoods.
For various complex networks (\eg bipartite networks~\cite{gao2018bine} and signed networks~\cite{yuan2017sne}), researchers have also proposed a variety of embedding methods by adapting the random walk methods.

Recently, Graph neural networks (GNNs) have received tremendous attention due to the power in learning effective representations for graphs~\cite{xugraph, xu2018graph}. 
Most GNNs can be summarized as a message-passing scheme where the node representations are updated by aggregating and transforming the information from the neighborhood~\citep{gilmer2017neural}.
GNNs have a partial intersection but use the deep learning methods instead of matrix factorization and random walk and can better describe the network structure~\citep{wu2019comprehensive}. 
A lot of GNN models show a better performance than the shadow lookup embeddings\citepp{kipf2016semi,velickovic2017graph,hamilton2017inductive}.
Most GNNs are designed for unsigned social networks whose links are only positive.
For signed networks, some signed GNNs~\cite{derr2018signed, huang2019signed} are proposed to model the balance theory using convolution or attention mechanism.
But they cannot handle the signed bipartite networks, because there are no links between nodes in the same sets.
It is not trivial to transfer these models to signed bipartite networks.

\section{Balance Theory in Signed Bipartite Networks}
\label{sec:balance_thoery}

\begin{figure*}
    \centering
    \includegraphics[width=\textwidth]{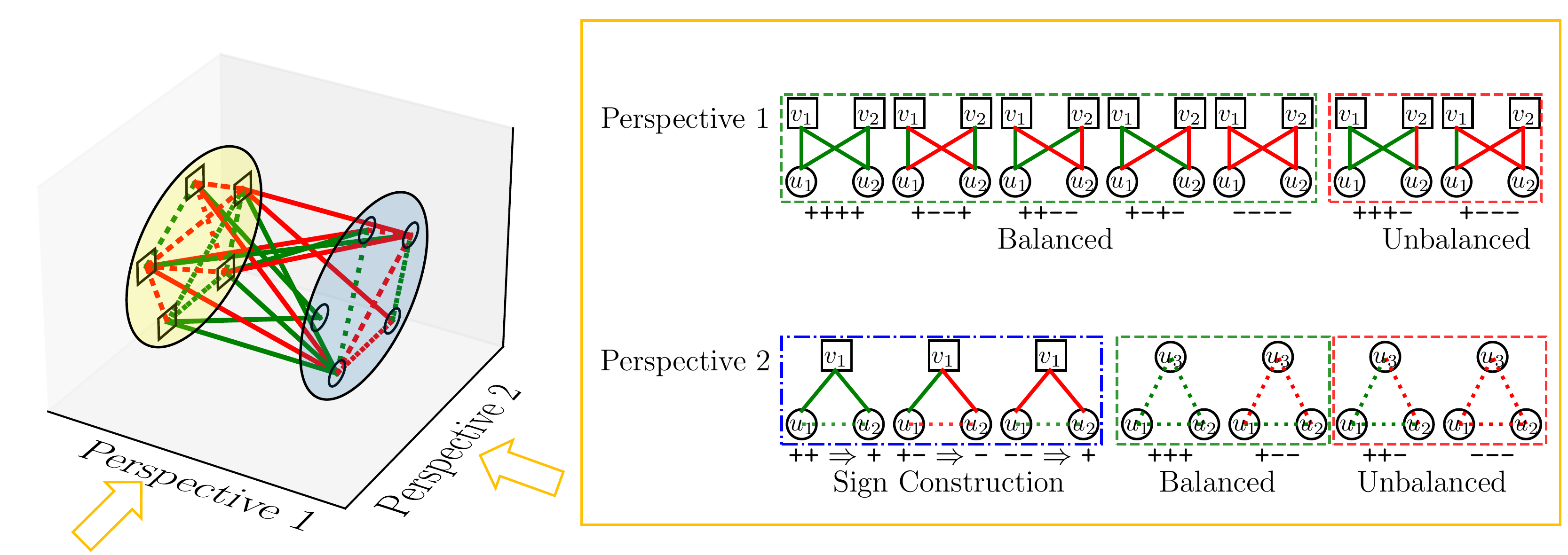}
    \caption{For a signed bipartite network, there exist two different analysis perspectives. Perspective 1 offers to analyze the signed butterfly isomorphism.  For Perspective 2, we can analyze the signed triangle isomorphism by sign construction. }
    \label{fig:perspective}
\end{figure*}

For signed networks, balance theory is one of the most fundamentally studied social theories, which is originated in social psychology in the 1950s~\cite{heider1944social}. 
It discusses that due to the stress or psychological dissonance, people will strive to minimize the unbalanced state in their personal relationships, and hence they will change to balanced social settings.
Specifically, triads with an even number of negative edges are defined as balanced.
However, previous researches on balance theory are focused on unbipartite signed networks, measuring balance theory in signed bipartite networks is less studied.
In this section, we give two perspectives to analyze balance theory in signed bipartite networks.

\subsection{Signed Bipartite Networks}
\label{sec:signed_bipratite_networks}

\begin{table}
\centering
\caption{Statistics on Signed Bipartite Networks.}
\label{tab:dataset}
    \setlength{\tabcolsep}{0.8mm}{
\begin{tabular}{lccccc}
\toprule
 & Bonanza
 &  \begin{tabular}[c]{@{}c@{}}U.S. \\House \end{tabular} 
 &  \begin{tabular}[c]{@{}c@{}}U.S. \\Senate \end{tabular}  
 &  \begin{tabular}[c]{@{}c@{}}Preliminary \\Review \end{tabular}  
 & \begin{tabular}[c]{@{}c@{}}Final \\Review \end{tabular} \\
\midrule
$|\mathcal{U}|$ & 7,919 & 515 & 145 & 182 & 182 \\
$|\mathcal{V}|$ & 1,973 & 1,281 & 1,056 & 304 & 304 \\
$|\E| = |\E^+| + |\E^-|$& 36,543 & 114,378 & 27,083 & 1,170 & 1,170 \\
\%  Positive Links  & 0.980 & 0.540 & 0.553 & 0.403 & 0.397 \\
\%  Negative Links & 0.020 & 0.460 & 0.447 & 0.597 & 0.603 \\
\midrule
\end{tabular}
}
\end{table}

Firstly, we describe our datasets used in this paper.
The first dataset is from the e-commerce website Bonanza\footnote{https://www.bonanza.com/}, which is similar to eBay\footnote{https://www.ebay.com/} or Taobao\footnote{https://www.taobao.com/}. 
Users can purchase products from a seller and rate the seller with ``Positive'', ``Neutral'', or ``Negative'' scores.
In this dataset, $\mathcal{U}$ represents the buyers, and $\mathcal{V}$ represents the sellers.
The next two datasets (\ie U.S. Senate and U.S. House) are from the 1st to 10th United States Congress vote records.
It is collected from the Govtrack.us\footnote{https://www.govtrack.us/}.
The senators or representatives $\mathcal{U}$ can vote ``Yea'' or ``Nay'' for bills $\mathcal{V}$, which is the positive or negative links respectively.
The above datasets are collected and used by \cite{derr2019balance}\footnote{https://github.com/tylersnetwork/signed\_bipartite\_networks}.
The last dataset is the peer review data from a top computer science conference\footnote{Due to anonymity, we removed the name of the conference.}. 
Reviewers $\mathcal{U}$ can give ``SA'' (Strong Accept), ``A'' (Accept), ``WA'' (Weak Accept), ``WR'' (Weak Reject), ``R'' (Reject), and ``SR'' (Strong Reject) to papers $\mathcal{V}$ after reviewing.
We regard ``SA'', ``A'', and ``WA'' as positive links and ``SR'', ``R'' and ``WR'' as negative links.
It’s worth mentioning that in most computer science conference peer reviews, there is usually a rebuttal phase when authors can point out errors in the reviews and help clarify reviewers' misunderstandings.
It has proven to play a critical role in the final decision made by the meta-reviewers and the reviewers\footnote{https://icml.cc/FAQ/AuthorResponse}. 
Besides, during the rebuttal phase, the reviewers can see the scores of other reviewers and make adjustments to their review comments and scores based on the author’s response and other reviewers' comments.
Therefore, the peer review dataset is divided into two parts: Preliminary Review and Final Review.

We list the statistics of datasets in \tableref{tab:dataset}.
From \tableref{tab:dataset}, we can find that in different scenarios, the negative ratio varies.
In the scenario of product reviews, the ratio of negative links is relatively lower (\ie 0.02).
Buyers rarely give bad rates to sellers.
In the scenario of bill vote, the proportion of negative links increases comparing to the scenario of product reviews (\ie 0.460 and 0.447).
In many bills, it is more difficult for legislators to reach consensus due to different political standpoints.
In the scenario of peer reviews, the ratio of negative links is higher than the ratios of positive links (\ie $0.603 > 0.397$).
In the top conferences of computer science, the acceptance rate needs to be controlled (\eg about 20\% in CIKM\footnote{https://www.openresearch.org/wiki/CIKM}), so reviewers usually raise their standards for reviewing the paper, which will have a greater probability of giving negative reviews.
Surprisingly, after the rebuttal phase, the proportion of negative links has slightly risen (\ie from 0.597 to 0.603).

\subsection{Signed Caterpillars and Signed Butterflies}
The ``butterfly'' is the most basic motif that models cohesion in an unsigned bipartite network, which is the complete 2×2 biclique~\cite{sanei2018butterfly}.
Base on the butterfly definition, \cite{derr2019balance} extends it to the signed butterfly by giving signs to the links in classical butterfly isomorphism.
Except for signed butterfly definition, \cite{derr2019balance} denotes ``signed caterpillars'' as paths of length that are missing just one link to becoming a signed butterfly.
They use signed butterflies to investigate balance theory in signed bipartite networks.

According to the definition of ~\cite{derr2019balance}, we use the notation $\bigcirc  ~\bigcirc~\bigcirc~\bigcirc$ to denote a signed butterfly isomorphism class that represents the links between $\mathcal{U}$ and $\mathcal{V}$ (\ie $u_1\rightarrow^{\bigcirc} v_1, u_1\rightarrow^{\bigcirc} v_2, u_2\rightarrow^{\bigcirc} v_1, u_2\rightarrow^{\bigcirc} v_2$). 
Due to the symmetry of the structure, we can get 7 non-isomorphic signed butterflies classes.
We show them in Perspective 1 in \figref{fig:perspective} and divide them into two categories, balanced and unbalanced.
For example, isomorphism class $\texttt{+}\texttt{+}\texttt{+}\texttt{+}$ and  $\texttt{-}\texttt{-}\texttt{-}\texttt{-}$ denote the classes having all positive or all negative links, respectively.
In the scenario of peer reviews, we can interpret isomorphism class $\texttt{+}\texttt{+}\texttt{+}\texttt{+}$ as the situations where reviewer $u_1$ and reviewer $u_2$ both give ``Accept'' to paper $v_1$ and paper $v_2$ (\ie $u_1\rightarrow^+ v_1, u_1\rightarrow^+ v_2, u_2\rightarrow^+ v_1, u_2\rightarrow^+ v_2$).
Similarly, we can interpret isomorphism class $\texttt{-}\texttt{-}\texttt{-}\texttt{-}$ as reviewer $u_1$ and reviewer $u_2$ both reject paper $v_1$ and paper $v_2$ (\ie $u_1\rightarrow^- v_1, u_1\rightarrow^- v_2, u_2\rightarrow^- v_1, u_2\rightarrow^- v_2$).
Except isomorphism class $\texttt{+}\texttt{+}\texttt{+}\texttt{+}$ and $\texttt{-}\texttt{-}\texttt{-}\texttt{-}$, isomorphism class $\texttt{+}\texttt{+}\texttt{-}\texttt{-}$, $\texttt{+}\texttt{-}\texttt{+}\texttt{-}$, and $\texttt{-}\texttt{-}\texttt{-}\texttt{-}$ are balanced since they have an even number of negative links.
In fact, the definition of signed butterflies can be viewed as analyzing the signed bipartite network from Perspective 1.

\subsection{Signed Triangles in Signed Bipartite Networks}

For signed bipartite networks, the nodes of the same set are not connected.
Therefore, we propose a new sign construction process by judging the sign of the link from $\mathcal{U}$ to $\mathcal{V}$.
After sign construction, we have signed links between nodes in the same set.
It means that we  have two new signed networks for $\mathcal{U}$ and $\mathcal{V}$ after sign construction.

As shown in Perspective 2 in \figref{fig:perspective}, when $u_1$ and $u_2$ have links with same sign on $v_1$ (\ie, $u_1\rightarrow^{+} v_1, u_2\rightarrow^{+} v_1$ or $u_1\rightarrow^{-} v_1, u_2\rightarrow^{-} v_1$), we construct a positive links between $u_1$ and $u_2$ (\ie $\texttt{+}\texttt{+}\Rightarrow \texttt{+}$ and $\texttt{-}\texttt{-}\Rightarrow \texttt{+}$).
When $u_1$ and $u_2$ have different link signs on $v_1$ (\ie, $u_1\rightarrow^{+} v_1, u_2\rightarrow^{-} v_1,$), we construct a negative links between $u_1$ and $u_2$ (\ie $\texttt{+}\texttt{-}\Rightarrow \texttt{-}$).
Since $\mathcal{U}$ is a set of people nodes (\eg Buyer, Congress, and Reviewer), the positive and negative links can be regard as agreement and disagreements.
For $\mathcal{V}$, the positive link can be viewed as similarity and vice versa.
After constructing the sign links between nodes of the same types, we can use the balance theory analysis in the classical signed networks.
We can calculate the ratio of balanced triads (\ie Triads with an even number of negative edges) in all triads~\cite{leskovec2010signed}.
The signed triangles $\texttt{+}\texttt{+}\texttt{+}$ and $\texttt{+}\texttt{-}\texttt{-}$ are balanced as the principle that ``\textit{the friend of my friend is my friend, the enemy of my enemy is my friend}''.

\subsection{Balance Theory Analysis}
 
In this subsection, we analyze the balance theory in different datasets from different perspectives.
From Perspective 1, we follow \cite{derr2019balance} and calculate the percentage each isomorphism class takes up of the total signed butterfly count in each dataset as ``\%''.
Besides, we also calculate ``\%E" as the expectation of signed butterflies when randomly reassigning the positive and negative signs to the signed bipartite network. 
For example, ``\%E'' for the isomorphism class $\texttt{+}\texttt{-}\texttt{-}\texttt{-}$ is 
\[
\left(\begin{array}{c}
4 \\
1
\end{array}\right)\left(\left(\left|\mathcal{E}^{+}\right| /|\mathcal{E}|\right) \times\left(\left|\mathcal{E}^{-}\right| /|\mathcal{E}|\right)^{3}\right).
\]
For Perspective 2, we count the percentage of each signed triangles as ``\%'' and the expectation of such signed triangles as ``\%E".
Similarly, ``\%E'' in $\mathcal{U}$ for $\texttt{+}\texttt{-}\texttt{-}$ is 
\[
\left(\begin{array}{c}
3 \\
1
\end{array}\right)\left(\left(\left|\mathcal{E}^{+}_\mathcal{U}\right| /|\mathcal{E}_\mathcal{U}|\right) \times\left(\left|\mathcal{E}_\mathcal{U}^{-}\right| /|\mathcal{E}_\mathcal{U}|\right)^{2}\right),
\]
where $|\mathcal{E}| = |\mathcal{E}^-_\mathcal{U}| + |\mathcal{E}^+_\mathcal{U}|$, and $\mathcal{E}^+_\mathcal{U}$ and $\mathcal{E}^-_\mathcal{U}$ are the positive and negative edges in $\mathcal{U}$, respectively. 
Since $\mathcal{U}$ is the set of people nodes, which is easy to describe, we only list the results of $\mathcal{U}$. 
\begin{table*}[!htp]
\centering
\caption{Balance theory analysis on five real-world datasets from two perspectives.}
\label{tab:balance_analysis}
\begin{tabular}{lccccc}

\toprule
 & Bonanza
 &  \begin{tabular}[c]{@{}c@{}}U.S. \\House \end{tabular} 
 &  \begin{tabular}[c]{@{}c@{}}U.S. \\Senate \end{tabular}  
 &  \begin{tabular}[c]{@{}c@{}}Preliminary \\Review \end{tabular}  
 & \begin{tabular}[c]{@{}c@{}}Final \\Review \end{tabular} \\
\midrule

Signed Butterfly Isomorphism $\texttt{+}\texttt{+}\texttt{+}\texttt{+}$ (\%,  \%E) & (0.986, 0.922)	& (0.244, 0.085) & (0.262, 0.094) & (0.109, 0.026) & (0.115$\uparrow$, 0.025) \\
Signed Butterfly Isomorphism $\texttt{+}\texttt{-}\texttt{-}\texttt{+}$ (\%,  \%E) & (0.000, 0.001)	& (0.109, 0.123) & (0.108, 0.122) & (0.109, 0.116) & (0.072$\downarrow$, 0.115) \\
Signed Butterfly Isomorphism $\texttt{+}\texttt{+}\texttt{-}\texttt{-}$ (\%,  \%E) & (0.001, 0.001)	& (0.111, 0.123) & (0.110, 0.122) & (0.101, 0.116) & (0.057$\downarrow$, 0.115) \\
Signed Butterfly Isomorphism $\texttt{+}\texttt{-}\texttt{+}\texttt{-}$ (\%,  \%E) & (0.000, 0.001)	& (0.186, 0.123) & (0.184, 0.122) & (0.156, 0.116) & (0.215$\uparrow$, 0.115) \\
Signed Butterfly Isomorphism $\texttt{-}\texttt{-}\texttt{-}\texttt{-}$ (\%,  \%E) & (0.000, 0.000)	& (0.147, 0.045) & (0.133, 0.040) & (0.249, 0.127) & (0.315$\uparrow$, 0.133) \\

 Balanced Signed Butterfly Summary (\%,  \%E) & (\textbf{0.988}, 0.924) & (\textbf{0.798}, 0.500) & (\textbf{0.798}, 0.500) & (\textbf{0.724}, 0.501) & (\textbf{0.774$\uparrow$}, 0.501)\\
\midrule

Signed Butterfly Isomorphism $\texttt{+}\texttt{+}\texttt{+}\texttt{-}$ (\%,  \%E) &	(0.012, 0.076) &	(0.118, 0.289) & (0.122, 0.302) &	(0.070, 	0.156) & (0.075$\uparrow$, 0.151) \\
Signed Butterfly Isomorphism $\texttt{+}\texttt{-}\texttt{-}\texttt{-}$ (\%,  \%E) &	(0.000, 0.000) &	(0.085, 0.211) & (0.081, 0.197) &	(0.206, 	0.343) & (0.151$\downarrow$, 0.349) \\
Unbalanced Signed Butterfly Summary (\%,  \%E) & (\textbf{0.012}, 0.076)	& (\textbf{0.202}, 0.500) & (\textbf{0.202}, 0.500)	& (\textbf{0.276}, 0.499)	& (\textbf{0.226$\downarrow$}, 0.499)\\
\midrule
Signed Triangles Isomorphism $\texttt{+}\texttt{+}\texttt{+}$ in $\mathcal{U}$ (\%,  \%E) & 	(0.978, 0.949) &	(0.338, 0.217) &	(0.360, 0.248) &	(0.327, 0.213) &	(0.446$\uparrow$, 0.310) \\
Signed Triangles Isomorphism $\texttt{+}\texttt{-}\texttt{-}$ in $\mathcal{U}$ (\%,  \%E) & 	(0.011, 0.001) &	(0.476, 0.287) &	(0.436, 0.261) &	(0.451, 0.290) &	(0.346$\downarrow$, 0.212) \\
Balanced Signed Triangles Summary in $\mathcal{U}$  (\%,  \%E)  & 	(\textbf{0.989}, 0.950) & (\textbf{0.815}, 0.504) &	(\textbf{0.796}, 0.508) & (\textbf{0.778}, 0.504) & (\textbf{0.792$\uparrow$}, 0.522) \\
\midrule
Signed Triangle Isomorphism $\texttt{+}\texttt{+}\texttt{-}$ in $\mathcal{U}$ (\%,  \%E)  &  (0.011, 0.050) &	(0.176, 0.432) &	(0.189, 0.440) &	(0.194, 0.431) &	(0.195$\uparrow$, 0.444) \\
Signed Triangle Isomorphism $\texttt{-}\texttt{-}\texttt{-}$ in $\mathcal{U}$ (\%,  \%E)  & 	(0.000, 0.000) &	(0.009, 0.063) &	(0.015, 0.051) &	(0.027, 0.065) &	(0.012$\downarrow$, 0.034) \\
Unbalanced  Signed Triangles Summary in  $\mathcal{U}$(\%,  \%E)	 & (\textbf{0.011},  0.050)	& (\textbf{0.185}, 0.496) &	(\textbf{0.204}, 0.492) &	(\textbf{0.222}, 0.496) &	(\textbf{0.208$\downarrow$}, 0.478) \\

\bottomrule
\end{tabular}

\end{table*}

From \tableref{tab:balance_analysis}, we can find that the large majority of signed butterflies in signed bipartite networks are more balanced than expectation based on the link sign ratio in the given networks (\eg $0.988> 0.924$ in Bonanza).
For Perspective 2, signed triangles in signed networks are also more balanced than expectation (\eg $0.989 >0.950$ in Bonanza).
Although the perspectives are different, the conclusions are similar.
In the scenario of peer reviews, after rebuttal phase, the balance of signed bipartite networks increased (\ie $0.724 \rightarrow 0.774\uparrow$ and $0.778 \rightarrow 0.792\uparrow$).
It shows that through authors' feedback and reviewers' discussions, the reviewers' opinions have become more balanced, although ratio of the negative links increased.
From Perspective 1, the ratio of  isomorphism class $\texttt{+}\texttt{+}\texttt{+}\texttt{+}$, $\texttt{+}\texttt{-}\texttt{+}\texttt{-}$,
$\texttt{-}\texttt{-}\texttt{-}\texttt{-}$ increased (\ie $0.109 \rightarrow 0.115\uparrow$, $0.156 \rightarrow 0.215\uparrow$, and $0.239 \rightarrow 0.315\uparrow$), which means that reviewers made a more balanced adjustment to their review comments.
For Perspective 2, the ratio of signed triangles $\texttt{+}\texttt{+}\texttt{+}$ increased from $0.327$ to $0.446\uparrow$, which reflects that the reviewers are more balanced and consistent after the rebuttal phase.

\section{Problem Formulation}
\label{sec:problem_definiton}
In this section, we give the definition of \textsc{Link Sign Prediction} , which can be regarded as the main machine learning task for  signed bipartite networks.

Consider a signed bipartite network, $\mathcal{G} = (\mathcal{U}, \mathcal{V}, \E)$, 
where $\mathcal{U} = \{ u_1, u_2, ..., u_{|\mathcal{U}|} \}$ and  $\mathcal{V} = \{ v_1, v_2, ..., v_{|\mathcal{V}|}  \}$ represent two sets of  nodes with the number of nodes $|\mathcal{U}|$ and $|\mathcal{V}|$.
$\E \subset \mathcal{U} \times \mathcal{V}$ is the edges between $\mathcal{U}$ and $\mathcal{V}$.
$\E = \E^+ \bigcup \E^-$ is the set of edges between the two sets of nodes $\mathcal{U}$ and $\mathcal{V}$ where $\E^+ \bigcap \E^- = \varnothing$,  $\E^+$ and $\E^-$ represent the sets of positive and negative edges, respectively.
Given $\mathcal{G} = (\mathcal{U}, \mathcal{V}, \E)$, $u_{i}$ and $v_{j}$ from two different sets (their link sign is not observed), the goal is to find a mapping function $f(u_{i},v_{j}) \rightarrow \{-1, 1\}$.
For network embeddings methods or GNNs, it will learn the representation of the node $u_{i}$ and $v_{j}$ to get embeddings $z_{u_i} \in \mathbb{R}^{d_u}$ and $z_{v_j} \in \mathbb{R}^{d_v} $, and use the embeddings to get the results by $f(z_{u_i}, z_{v_j}) \rightarrow \{-1, 1\}$.

\section{Proposed Methodology}
\label{sec:model}

Based our discussion in \secref{sec:balance_thoery} and problem definition in \secref{sec:problem_definiton}, we proposed a new \underline{S}igned \underline{B}ipartite \underline{G}raph \underline{N}eural \underline{N}etworks (SBGNN) model to do \lsp task.

Vanilla GNNs usually follow a message passing scheme where the node representations are updated by aggregating and transforming the information from the neighborhood~\cite{you2020design}.

Specifically, for a graph $\mathcal{G} = (\V, \E)$, where $\V=\{v_1, v_2, ...v_{|V|}\}$ is the node set and $\E \subset \V \times \V$ is the edge set.
The goal of GNNs is to learn node representation $h_i$ for node $i$ based on an iterative aggregation of local neighborhoods.

\begin{figure*}
    \centering
    \includegraphics[width=\textwidth]{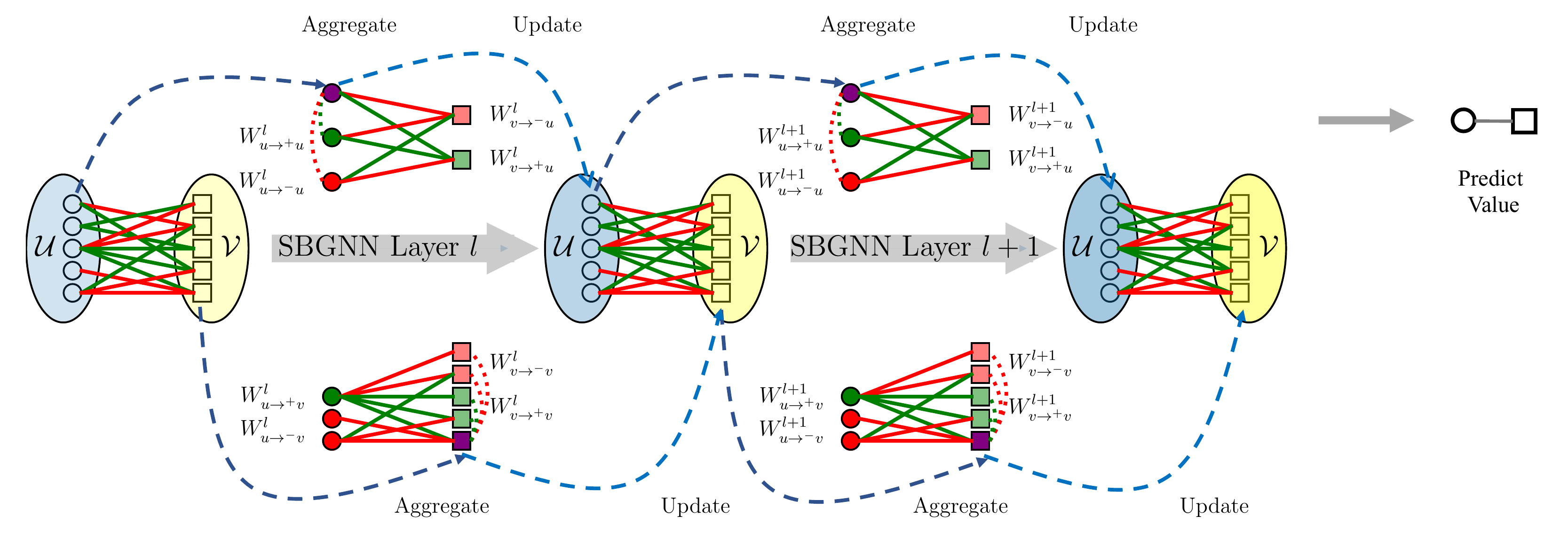}
    \caption{Illustration of SBGNN. SBGNN Layer includes Aggeregate and Update functions. 
    The aggregated message comes from the \textit{Set}$_1$ and \textit{Set}$_2$ with positive and negative links.
    After getting the embedding of the node $u_i$ and $v_i$, it can be used to predict the link sign relationship.}
    \label{fig:SBGNN}
\end{figure*}

For the $l$-th layer of a GNN, it can be written as:
\begin{equation}
 \begin{aligned}
     m_{j \rightarrow i} ^{(l)}(i, j) &= \textsc{Msg}^{(l)}\left(h_{i}^{(l)}, h_{j}^{(l)}, h_{e_{j i}}^{(l)}\right), j \in \mathcal{N}(j),\\
    m_{j \rightarrow i}^{(l)} (i)  &=\textsc{Agg}^{(l)}\left(\left\{m_{j \rightarrow i}^{(l)}(i, j) \mid j \in \mathcal{N}(i)\right\}\right),\\
    h_{i}^{(l+1)} &=\textsc{Upt}^{(l)}\left(h_{i}^{(l)}, m_{j \rightarrow i}^{(l)}(i) \right),
\end{aligned}
\end{equation}
where $\textsc{Msg}, \textsc{Agg}, \textsc{Upt}$ are functions for \textit{message construction}, \textit{message aggregation}, and \textit{vertex update function} at the $l$-th layer~\cite{li2020deepergcn}.
Most Graph Neural Networks~\cite{kipf2016semi} are designed for unsigned classical social networks.
They design the $\textsc{Msg}$ and $\textsc{Agg}$ function such as \textit{mean}~\cite{kipf2016semi}, \textit{max}~\cite{hamilton2017inductive},  \textit{sum}~\cite{xu2018powerful} or \textit{attention}~\cite{velickovic2017graph}.
In this paper, we follow the message passing scheme and show the SBGNN Layer in \figref{fig:SBGNN} including the design of $\textsc{Msg}$, $\textsc{Agg}$ and $\textsc{Upt}$.

\subsection{Message and Aggregation Function}
As we discussed in \secref{sec:balance_thoery}, comparing to traditional unsigned networks, the links in signed bipartite networks is cross-set and complex (\eg $u \rightarrow ^{+/-} v$).

The message function of vanilla GNNs cannot be directly applied to signed bipartite networks. 
As shown in \figref{fig:SBGNN}, we design a new message function to aggregate messages from different sets of neighborhoods .
We define that \textit{Set}$_1$ refers to the set of node with different types, and \textit{Set}$_2$ refers to the set of nodes with the same types.
Message from \textit{Set}$_1$ and \textit{Set}$_2$ can be viewed as the modeling of Perspective 1 and Perspective 2 in \secref{sec:balance_thoery}, respectively.

\subsubsection{Message from Set$_1$}
For \textit{Set}$_1$, the type of the set is different from the type of the current node, and because it is a signed network, its links include both positive and negative relationships.
Neighborhood nodes under positive and negative links have different semantic relations.
So we use $W_{v\rightarrow^+ u}$ and $W_{v\rightarrow^- u}$ to aggregate the message from $v$ to $u$ with positive and negative links and $W_{u\rightarrow^+ v} $ and $ W_{u\rightarrow^- v}$ to aggregate the message from $u$ to $v$ with positive and negative links.
For example, in \figref{fig:SBGNN}, for the purple circle $u_1$, the red square $v_2$ and the green square $v_4$ are the positive and negative neighborhoods, respectively.

At SBGNN Layer $l$-th layer, we  use  $W^l_{v\rightarrow^+ u}, W^l_{v\rightarrow^- u}$ to collect the message from $v_j$ to $u_i$ by 
\begin{equation}
    \begin{aligned}
        m^l_{v\rightarrow^+ u} (v_j, u_i) & = \textsc{Msg}(h^l_{u_i}, h^l_{v_j}) = W^l_{v\rightarrow^+ u} \cdot h^l_{v_j}, v_j\in \mathcal{N}_{v\rightarrow^+ u} ({u_i}), \\
        m^l_{v\rightarrow^- u} (v_j, u_i) & = \textsc{Msg}(h^l_{u_i}, h^l_{v_j}) = W^l_{v\rightarrow^- u} \cdot h^l_{v_j}, v_j\in \mathcal{N}_{v\rightarrow^- u} ({u_i}).
    \end{aligned}
\end{equation}
where $\mathcal{N}_{v\rightarrow^+ u} (u_i)$ and $\mathcal{N}_{v\rightarrow^- u} (u_i)$ are the neighborhood with positive and negative links to $u_i$.
Similarly, we use we  use  $W^l_{u\rightarrow^+ v}, W^l_{u\rightarrow^- v}$ to collect the message from 
\textit{Set}$_1$ for $v_i$:
\begin{equation}
    \begin{aligned}
        m^l_{u\rightarrow^+ v} (u_j, v_i) & = \textsc{Msg}(h^l_{v_i},h^l_{u_j})= W^l_{u\rightarrow^+ v} \cdot h^l_{u_j}, u_j\in \mathcal{N}_{u\rightarrow^+ v} (v_i), \\
        m^l_{u\rightarrow^- v} (u_j, v_i) & = \textsc{Msg}(h^l_{v_i},h^l_{u_j})= W^l_{u\rightarrow^+ v} \cdot h^l_{u_j}, u_j\in \mathcal{N}_{u\rightarrow^- v} (v_i).
    \end{aligned}
\end{equation}
where $\mathcal{N}_{u\rightarrow^+ v}(v_i)$ and $\mathcal{N}_{u\rightarrow^- v}(v_i)$ are the neighborhood with positive and negative links to $v_i$.

\subsubsection{Message from Set$_2$}
As we said before, \textit{Set}$_2$ is the node set of the same type.
However, there are no links between nodes of the same type, so we need to construct an sign link between nodes of the same type through sign construction in \secref{sec:balance_thoery}.

After sign construction, we can aggregate message for the positive and negative links from $u_j$ to $u_i$ with $W_{u\rightarrow^+ u}$ and $W_{u\rightarrow^- u}$, respectively and $v_j$ to $v_i$ with $W_{v\rightarrow^+ v}$ and $W_{v\rightarrow^- v}$, respectively.

In \figref{fig:SBGNN}, for the purple circle $u_1$, the green circle $u_3$ and the red circle $u_5$ are the positive and negative neighborhoods because of the link $u_1\rightarrow^- v_2 \rightarrow^{-} u_3$ and $u_1\rightarrow^- v_2 \rightarrow^{+} u_5$.
For the purple square $v_5$, $v_1$ and $v_2$ are the negative neighborhoods, $v_3$ and $v_4$ are the positive neighborhoods based on our sign construction.

In summary, we can definite the message from \textit{Set}$_2$ as follows: 
\begin{equation}
    \begin{aligned}
        m^l_{u\rightarrow^+ u}(u_j, u_i) & = \textsc{Msg}(h^l_{u_i}, h^l_{u_j})= W^l_{u\rightarrow^+ u} \cdot h^l_{u_j}, u_j\in \mathcal{N}_{u\rightarrow^+ u} (u_i), \\
        m^l_{u\rightarrow^- u}(u_j, u_i) & = \textsc{Msg}(h^l_{u_i}, h^l_{u_j})= W^l_{u\rightarrow^- u} \cdot h^l_{u_j}, u_j\in \mathcal{N}_{u\rightarrow^- u} (u_i),
        \\
        m^l_{v\rightarrow^+ v} (v_j, v_i) & = \textsc{Msg}(h^l_{v_i}, h^l_{v_j})= W^l_{v\rightarrow^+ v} \cdot h^l_{v_j}, v_j\in \mathcal{N}_{v\rightarrow^+ v} (v_i), \\
        m^l_{v\rightarrow^- v} (v_j, v_i) & = \textsc{Msg}(h^l_{v_i}, h^l_{v_j})= W^l_{v\rightarrow^- v} \cdot h^l_{v_j}, v_j\in \mathcal{N}_{v\rightarrow^- v} (v_i),
    \end{aligned}
\end{equation}
where $\mathcal{N}_{u\rightarrow^- u} (u_i)$, $\mathcal{N}_{u\rightarrow^- u}(u_i)$, $\mathcal{N}_{v\rightarrow^+ v}(v_i)$, and $\mathcal{N}_{v\rightarrow^- v}(v_i)$ are positive and negative neighborhoods for $u_i$ and $v_i$, respectively.

\subsubsection{Aggregation Design}

For \textit{message aggregation}, it is commonly a differentiable, permutation
invariant set function (\eg \textit{mean}, \textit{max},  and \textit{sum}) that take a countable message set $\{m_{v}(u_i)| u_i\in \mathcal{N}(v) \}$ as input; and output a message vector $m_v$.
In this paper, we use mean ($\textsc{Mean}$) and graph attention ($\textsc{Gat}$) aggregation functions.

For the $\textsc{Mean}$ aggregation function, we get $m^l (u_i)$ and $m^l(v_i)$ by $\textsc{Mean}$ the message of neighborhoods:

\begin{equation}
\begin{aligned}
  m_{ {\leadsto} u  }^{l} (u_i) &= \textsc{Mean} (\{ m^l_{\leadsto u}(j) , \forall j \in \mathcal{N}_{\leadsto u}(u_i) \}), \\ 
  m_{ {\leadsto} v }^{l} (v_i) &= \textsc{Mean} (\{ m^l_{\leadsto v}(j) , \forall j \in \mathcal{N}_{\leadsto v}(v_i) \})
    \end{aligned}
\end{equation}
where $\leadsto u$ is a relationship of links to $u$  (\eg, $v\rightarrow^{+} u$) and $\leadsto v$ is a relationship of links to $v$  (\eg, $u\rightarrow^{+} v$).

For a graph attention function, it will firstly compute $\alpha^{i j}_{\leadsto}$ for node $i$ and node $j$ by the attention mechanism $\vec{\bf a}_{ \leadsto}$ and LeakyReLU nonlinearity activation function (with negative input slope $\alpha$ = 0.2) as:
   \begin{equation}
    \begin{aligned}
      \label{eq:gat1}
      \alpha^{ij}_{ \leadsto}  = \frac{\exp\left(\text{LeakyReLU}\left(\vec{\bf a}_{ \leadsto}^\top \cdot [{\bf W}^l_{ \leadsto}{h^l_i}\|{\bf W}^l_{ \leadsto}{h^l_j}]\right)\right)}{\sum_{k\in\mathcal{N}_{ \leadsto}(i)} \exp\left(\text{LeakyReLU}\left(\vec{\bf a}_{ \leadsto}^\top \cdot [{\bf W}^l_{ \leadsto}{h^l_i}\|{\bf W}^l_{ \leadsto}{h^l_k}]\right)\right)}, \\
    \end{aligned}
  \end{equation}
 where $\|$ is is the concatenation operation, $\cdot^{\top}$ represents transposition, $\mathcal{N}_{ \leadsto}(i)$ is the neighborhoods of node $i$ under the definition of $\leadsto$ (\eg, $u\rightarrow^{+} v$) , ${\bf W}^l$ is the weight matrix parameter.
Then we can compute $m^l(u_i)$ and $m^l(v_i)$ with $\alpha_{\leadsto}$:
\begin{equation}
    \begin{aligned}
  m_{ {\leadsto} u  }^{l} (u_i) &= \sum_{j \in \mathcal{N}_{\leadsto u}(u_i) } \alpha^{u_i j}_{\leadsto u} \cdot m^l_{\leadsto u}(j), \\ 
  m_{ {\leadsto} v  }^{l} (v_i) &= \sum_{j \in \mathcal{N}_{\leadsto v}(v_i) } \alpha^{v_i j}_{\leadsto v} \cdot m^l_{\leadsto v}(j),
    \end{aligned}
\end{equation}
where $\leadsto u$ is a relationship of links to $u$  (\eg, $v\rightarrow^{+} u$) and $\leadsto v$ is a relationship of links to $v$  (\eg, $u\rightarrow^{+} v$).
The attention aggregation can be seen as a learnable weighted average function.

\subsection{Update Function}
For our \textit{vertex update function}, we concatenate the messages from different neighborhoods with origin node features and apply it to an $\textsc{Mlp}$ to get the final node representation:
\begin{equation}
    \begin{aligned}
      h^{l+1}_u & = \textsc{Mlp}(h^l_u ~\|~  m^l_{v\rightarrow^+ u} ~\|~  m^l_{v\rightarrow^- u} ~\|~  m^l_{u\rightarrow^+ u} ~\|~  m^l_{u\rightarrow^- u}), \\
      h^{l+1}_v & = \textsc{Mlp}(h^l_v ~\|~  m^l_{u\rightarrow^+ v} ~\|~  m^l_{u\rightarrow^- v} ~\|~  m^l_{v\rightarrow^+ v} ~\|~  m^l_{v\rightarrow^- v}),
    \end{aligned}
\end{equation}
where $\|$ is the concatenation operation. 
More specificlly, the $\textsc{Mlp}$is a two-layer neural networks with $\textsc{Dropout}$ layer and $\textsc{Act}$ activation function:
\begin{equation}
    \textsc{Mlp}(x) = W_2\Big(\textsc{Act}\big(\textsc{Dropout} (W_1x + b_1)\big)\Big) + b_2,
\end{equation}
where $W_1, b_1$ and $W_2, b_2$ is the parameters for this $\textsc{MLP}$, and $\textsc{Dropout}$ is the dropout function ($p=0.5$ in this paper) and $\textsc{Act}$ is the activation function (\eg $\mathrm{PReLU}$~\cite{he2015delving} in this paper) .

\subsection{Loss Function}
\label{sec:loss_function}
After getting embeddings $z_{u_i} \in \mathbb{R}^{d_u}$ and $z_{v_j} \in \mathbb{R}^{d_v}$  of the node $u_{i}$ and $v_{j}$,  we can use following methods to get the prediction value for $u_i\rightarrow v_j$.
The first one is the product operation:
\begin{equation}
    y_{pred} = \mathrm{sigmoid} (z_{u_i}^{\top} \cdot z_{v_j}),
\end{equation}
where $\cdot^\top$ is the transpose function and $\mathrm{sigmoid}$ is the sigmoid function $f(x) = \frac{1}{1+e ^{-x}}$.
This method should keep the embedding dimension the same (\ie $d_u = d_v$).
Another method is to use an \textsc{Mlp} to predict the values by
\begin{equation}
    y_{pred} =  \mathrm{sigmoid}\big(\textsc{Mlp}( z_{u_i} ~\|~ z_{v_i} )\big)
\end{equation}
where $\textsc{Mlp}$ is a two layer neural networks, $\|$ is the concatenation operation.
The $\textsc{Mlp}$ can be viewd as the Edge Learner in~\cite{agrawal2019learning}.

After getting the prediction values, we use binary cross entropy as the loss function:
\begin{equation}
     \quad
\mathcal{L} = - w \left[ y \cdot \log y_{pred} + (1 - y) \cdot \log (1 - y_{pred}) \right]
\end{equation}
where $w$ is the rescaling weight for the unblanced negative ratios (It is the weights inversely proportional to class frequencies in the input data); $y$ is the ground truth with mapping $\{-1, 1\}$ to $\{0, 1\}$.

\subsection{Training SBGNN}

With the design of our SBGNN model, the training procedure is summarized in \algref{alg:algorithm1}. 
\begin{algorithm}
  \caption{\name Algorithm}
  \label{alg:algorithm1}
  \begin{algorithmic}[1]
    \renewcommand{\algorithmicrequire}{\textbf{Input:}}
    \renewcommand{\algorithmicensure}{\textbf{Output:}}
    \REQUIRE {
      Signed Bipartite Graph $\mathcal{G}(\mathcal{U}, \mathcal{V}, \E)$;\\
      Encoder Aggregators $Enc$;\\
      SBGNN Layer Number $L$;\\
      Epoch $T$;
    }
    \ENSURE{
      Node representation $Z_{\mathcal{U}}, Z_{\mathcal{V}}$
    }
    \\
    \STATE{Prepare original node embeddings $z^0_u, z^0_v, \forall u \in \mathcal{U}, \forall v \in \mathcal{V}$.}
    \STATE{Initialize the parameters of SBGNN model.}
    
    \FOR{$epoch=1,...,T$}
    \STATE{Get neighborhoods $\mathcal{N}_{v\rightarrow^+ u}(u_i)$, $\mathcal{N}_{v\rightarrow^- u}(u_i)$,
     $\mathcal{N}_{u\rightarrow^+ u}(u_i)$,
     $\mathcal{N}_{u\rightarrow^- u}(u_i)$,
     $ \forall u_i \in \mathcal{U}$}
     \STATE{Get neighborhoods $\mathcal{N}_{u\rightarrow^+ v}(v_i)$, $\mathcal{N}_{u\rightarrow^- v}(v_i)$,
     $\mathcal{N}_{v\rightarrow^+ v}(v_i)$,
     $\mathcal{N}_{v\rightarrow^- v}(v_i)$,
     $ \forall v_i \in \mathcal{V}$}
    \FOR{$l=0...L-1$}
   
    \STATE{
      $z_{u_i}^{l+1} \leftarrow Enc^l\Big(h^l_{u_i} ,  m^l_{v\rightarrow^+ u}(u_i) ,  m^l_{v\rightarrow^- u}(u_i)$, 
      \\ $m^l_{u\rightarrow^+ u}(u_i) ,  m^l_{u\rightarrow^- u}(u_i) \Big)$,  $\forall u_i \in \mathcal{U}$
    }
    
    \STATE{
      $z_{v_i}^{l+1} \leftarrow Enc^l\Big(h^l_{v_i} ,  m^l_{u\rightarrow^+ v}({v_i}) ,  m^l_{u\rightarrow^- v}({v_i})$,
      \\  $m^l_{v\rightarrow^+ v}({v_i}) ,  m^l_{v\rightarrow^- v}({v_i})\Big)$, $\forall v_i \in \mathcal{V}$
    }
    
    \ENDFOR
    \STATE{Compute loss $\sum\limits_{u \rightarrow v \in \E} \mathcal{L}_{loss}(u \rightarrow v)$ with $Z_{\mathcal{U}}$ and $Z_{\mathcal{V}}$}
    \STATE{Back propagation, update parameters.}
    \ENDFOR
    \RETURN {$Z^l_{\mathcal{U}}, Z^l_{\mathcal{V}}$}
  \end{algorithmic}
\end{algorithm}

\algref{alg:algorithm1} demonstrates that our SBGNN is a layer-by-layer architecture design, where $Enc$ can be any powerful GNN aggregators like $\textsc{Mean}$ or $\textsc{Gat}$.

\section{Experiments}
\label{sec:experiments}
In this section, we evaluate the performance of our proposed SBGNN on real-world datasets.
We first introduce the datasets, baselines, and metrics for experiments, then present the experimental results of SBGNN and baselines.
Finally, we analyze our models from parameter analysis and ablation study. 

\subsection{Experimental Settings}

\subsubsection{Datasets}

As previously discussed in \secref{sec:signed_bipratite_networks}, we choose four datasets for this study, namely, Bonanza, U.S. House, U.S. Senate, and Review (we use Final Review as the Review dataset).
Following the experimental settings in \cite{derr2019balance}, we  randomly select 10\% of the links as test set, utilize a random 5\% for validation set, and the remaining 85\% as training set for each of our datasets.
We run with different train-val-test splits for 5 times to get the average scores.

\subsubsection{Competitors}
We compare our method SBGNN with several baselines including Random Embeddings, Unsigned Network Embeddings, Signed/Bipartite Network Embeddings, and Signed Butterfly Based Methods as follows.

\vpara{Random Embeddings: }
It generates $d$ dimensional random values from a uniform distribution over $[0, 1)$ (\ie $z=(z_1, z_2, ..., z_d)$, $z_i \in [0, 1)$). 
Given embeddings $z_{u_i}$ and $z_{v_j}$, we concatenate them and use a Logistic Regression (\textsc{Lr}) to predict the value of $u_i$ and $v_i$.
\textsc{Lr} will be trained on the training set, and make predictions on the test set.
Since \textsc{Lr} has learnable parameters, this method can be viewed as the lower bound of the graph representation learning methods.

\vpara{Unsigned Network Embeddings: }
Comparing random embeddings, we use some classical unsigned network embedding methods (\eg DeepWalk~\cite{perozzi2014deepwalk}\footnote{https://github.com/phanein/deepwalk}, Node2vec~\cite{grover2016node2vec}\footnote{https://github.com/aditya-grover/node2vec}, LINE~\cite{tang2015line}\footnote{https://github.com/tangjianpku/LINE}).
By keeping only positive links, we input unsigned networks to such unsigned network embedding methods to get embeddings for $u_i$ and $v_j$.
As same as Random Embeddings, we concatenate embeddings $z_{u_i}$ and $z_{v_j}$, and use \textsc{Lr} to predict the sign of links.

\vpara{Signed/Bipartite Network Embedding: }
We use Signed or/and Bipartite Network Embedding as our baselines.
More specifically, we use SiNE~\cite{wang2017signed}\footnote{
http://www.public.asu.edu/\textasciitilde swang187/codes/SiNE.zip} to learn the embeddings for $u_i$ and $v_i$ after sign link construction in \secref{sec:balance_thoery}.
For BiNE~\cite{gao2018bine}\footnote{https://github.com/clhchtcjj/BiNE}, we remove the negative links between $\mathcal{U}$ and $\mathcal{V}$ .
We use BiNE to get embeddings $z_{u_i}$ and $z_{v_j}$ and use \textsc{Lr} to predict the sign of links with concatenating $z_{u_i}$ and $z_{v_j}$.
Compared with the unsigned network embeddings methods, we try to let the representation learn the structural information (\eg links between $\mathcal{U}$ and $\mathcal{V}$ and the link sign) instead of just relying on the downstream classifier.
SBiNE~\cite{zhang2020sbine} is a representation learning method for signed bipartite networks, which preserves the first-order and second-order proximity.
Instead of a two-stage model, SBiNE uses single neural networks with sigmoid nonlinearity function to predict the value of $u_i$ and $v_j$.

\vpara{Signed Butterfly Based Methods:}
Based on the analysis of signed butterfly isomorphism, \cite{derr2019balance} proposes a variety of methods for \lsp, including SCsc, MFwBT, and SBRW \footnote{https://github.com/tylersnetwork/signed\_bipartite\_network}.
Specifically, SCsc is a balance theory guided feature extraction method.
MFwBT is the matrix factorization model with balance theory.
SBRW is the signed bipartitle random  walk method.
Due to the findings about receiving aid in prediction from balance theory will always perform better than the methods that only use generic signed network information ~\cite{derr2019balance}, we take the SCsc as the most competitive baseline for our SBGNN model.

\vpara{Signed Bipartite Graph Neural Networks:}
For our SBGNN, we try two different aggregation design (\ie $\textsc{Mean}$ and $\textsc{Gat}$) and remark it as SBGNN-$\textsc{Mean}$ and SBGNN-$\textsc{Gat}$, respectively.

For a fair comparison, we set all the node embedding dimension to 32  which is as same as that in SBiNE~\cite{zhang2020sbine} for all embedding based methods.
For other parameters in baselines, we follow the recommended settings in their original papers.
For embedding methods, we use the balanced class weighted Logistic Regression in Scikit-learn~\cite{scikitlearn}\footnote{https://scikit-learn.org/stable/index.html}.
For SBiNE, we use PyTorch~\cite{paszke2019pytorch} to implement it by ourselves.
For our SBGNN, we also use PyTorch to implement our model.
We use Adam optimizer with an initial learning rate of 0.005 and a weight decay of 1e-5.
We run 2000 epochs for SBGNN and choose the model that performs the best AUC metrics on the validation set.

\subsubsection{Evaluation Metrics}
Since \lsp is a binary classification problem, we use AUC, Binary-F1, Macro-F1, and Micro-F1 to evaluate the results.
These metrics are widely used in \lsp~\cite{chen2018bridge,huang2021sdgnn}.
Note that, among all these four evaluation metrics, the greater
the value is, indicating the better the performance of the corresponding method.

\subsection{Experiment Results}
\label{sec:experiments-res}

\begin{table*}[!ht]
  \centering 

    \caption{The results of \lsp on four datasets. 
    Results not available are marked as `\#N/A'. 
    Two-tailed t-tests demonstrate the improvements of our SBGNN to the baseline SCsc are statistically significant ( $^*$ indicates p-value $\leq$ 0.05).}
\label{tb:experiment-result}
    \scalebox{0.92}{
    \setlength{\tabcolsep}{1.0mm}{
      \begin{tabular}{c|c|c|ccc|ccc|ccc|cc}
    \toprule
        \multicolumn{2}{c|}{} & \multicolumn{1}{c|}{\begin{tabular}[c]{@{}c@{}}Random \\ Embedding\end{tabular}}  
    & \multicolumn{3}{c|}{\begin{tabular}[c]{@{}c@{}}Unsigned \\Network Embedding\end{tabular}} 
    & \multicolumn{3}{c|}{\begin{tabular}[c]{@{}c@{}}Signed/Bipartite \\Network Embedding\end{tabular}} 
    & \multicolumn{3}{c|}{\begin{tabular}[c]{@{}c@{}}Signed Butterfly\\ Based Methods\end{tabular}}
    & \multicolumn{2}{c}{\begin{tabular}[c]{@{}c@{}}Signed Bipartite  \\Graph Neural Networks\end{tabular} } \\
    \midrule
   Dataset                     & Metric   & Random & Deepwalk & Node2vec & LINE & SiNE  & BiNE  & SBiNE &  SCsc  & MFwBT & SBRW & SBGNN-$\textsc{Mean}$ & SBGNN-$\textsc{Gat}$ \\
    \midrule
\multirow{4}{*}{Bonanza} 
& AUC & 0.5222 & 0.6176 & \underline{0.6185} & 0.6124 & 0.6088 & 0.6026 & 0.5525 &  \textbf{0.6524} & 0.5769 & 0.5315 & 0.5841 & 0.5769 \\
& Binary-F1 & 0.7282 & 0.7843 & 0.7530 & 0.6974  & 0.9557 & 0.7426 & 0.8514 &  0.6439 & 0.8927 & \textbf{0.9823}   & 0.9488$^*$ & \underline{0.9616}$^*$ \\
& Macro-F1 & 0.3868 & 0.4258 & 0.4087 & 0.3790 & \textbf{0.5422} & 0.4016 & 0.4538 &  0.3543 & 0.4813 & 0.5353 &  0.5311$^*$ & \underline{0.5404}$^*$ \\
& Micro-F1 & 0.5770 & 0.6497 & 0.6093 & 0.5424 & 0.9157 & 0.5960 & 0.7436 &  0.4843 & 0.8076 & \textbf{0.9652}  & 0.9044$^*$ &   \underline{0.9269}$^*$ \\
                        
    \midrule
\multirow{4}{*}{Review}   
& AUC & 0.5489 & 0.6324 & 0.6472 & 0.6236 & 0.5741 & \#N/A & 0.5329 &  0.5522 & 0.4727 & 0.5837  & \underline{0.6584}$^*$ & \textbf{0.6747}$^*$  \\
& Binary-F1 & 0.4996 & 0.5932 & 0.6141 & 0.5974  & 0.5247 & \#N/A & 0.4232 &  0.3361 & 0.4346 & 0.5423 & \underline{0.6128}$^*$ & \textbf{0.6366}$^*$ \\
& Macro-F1 & 0.5426 & 0.6268 & 0.6400 & 0.6120  & 0.5688 & \#N/A & 0.5262 &  0.4823 & 0.4696 & 0.5767   & \underline{0.6556}$^*$ &  \textbf{0.6629}$^*$ \\
& Micro-F1 & 0.5487 & 0.6325 & 0.6444 & 0.6137 & 0.5744 & \#N/A & 0.5521 &  0.5812 & 0.4752 & 0.5812 & \underline{0.6632}$^*$ & \textbf{0.6667}$^*$   \\

    \midrule
\multirow{4}{*}{U.S. House} 

& AUC & 0.5245 & 0.6223 & 0.6168 & 0.5892 & 0.6006 & 0.6103 & 0.8328  & 0.8274 & 0.8097 & 0.8224 & \underline{0.8474}$^*$ & \textbf{0.8481}$^*$   \\
& Binary-F1 & 0.5431 & 0.6401 & 0.6323  & 0.6304 & 0.6118 & 0.6068  & 0.8434 & 0.8375 & 0.8234 & 0.8335 & \underline{0.8549}$^*$ & \textbf{0.8560}$^*$    \\
& Macro-F1 & 0.5238 & 0.6215 & 0.6158 & 0.5883 & 0.5991 & 0.6097  & 0.8323 & 0.8267 & 0.8096 & 0.8219  & \underline{0.8463}$^*$ & \textbf{0.8471}$^*$   \\
& Micro-F1 & 0.5246 & 0.6224 & 0.6166 & 0.5892 & 0.5996 & 0.6108  & 0.8330 & 0.8274 & 0.8106 & 0.8226  & \underline{0.8468}$^*$ & \textbf{0.8476}$^*$   \\

    \midrule
\multirow{4}{*}{U.S. Senate} 

& AUC & 0.5251 & 0.6334 & 0.6260 & 0.5743 & 0.5875  & 0.6071 &  0.7998 &0.8163  & 0.7857 & 0.8142 & \underline{0.8209}$^*$  & \textbf{0.8246}$^*$  \\
& Binary-F1 & 0.5502 & 0.6603 & 0.6526 & 0.6159 & 0.5923 & 0.5968 & 0.8175 &  \underline{0.8294} & 0.8043 & 0.8291 & 0.8277  & \textbf{0.8320}  \\
& Macro-F1 & 0.5239 & 0.6325 & 0.6251 & 0.5722  & 0.5842 & 0.6037 & 0.7992 & 0.8148 & 0.7850 & 0.8131 & \underline{0.8177}$^*$  & \textbf{0.8215}$^*$  \\
& Micro-F1 & 0.5254 & 0.6347 & 0.6271 & 0.5732 & 0.5848 & 0.6042  & 0.8009 &  0.8160 & 0.7867 & 0.8145 & \underline{0.8183}$^*$  & \textbf{0.8221}$^*$  \\

    \bottomrule
\end{tabular}
}
}

\end{table*}

We show the results in \tableref{tb:experiment-result}.
We have bolded the highest value of each row and underlined the second value.
From \tableref{tb:experiment-result}, we make the following observations:

\begin{itemize}[leftmargin=*]
\item Even with random embedding, \textsc{Lr} can still achieve a certain effect on \lsp (\ie AUC > 0.5).

    It demonstrates that the downstream classifier has a certain predictive ability for \lsp when it is regarded as a two-stage model.

    \item After using network embeddings, the graph structure data is modeled into the node representation, which improves the prediction results.
    Even the worst-performing LINE outperforms random embedding by 17.3\%, 13.6\%, 12.3\%, and 9.4\% on  AUC in Bonanza, Review, U.S. House, and U.S. Senate, respectively.
    It demonstrate that graph structure information is helpful for \lsp.
    In unsigned network embedding methods, Node2vec has made the best results in the unsigned network embeddings methods.
    We guess that the biased random walk mechanism may be able to explore the graph structure more effectively.
    \item For signed or bipartite model (\ie SiNE and BiNE), the information for sign and bipartite network structure can contribute to the node representation learning.
    SiNE is more effective than BiNE (\eg  AUC in Bonanza (0.6088 > 0.6026)), indicating that link sign information may be more important than the link relationship.
    The performance of SBiNE is not as good as that in \cite{zhang2020sbine}.
    This can be due to the fact of different data splits and implementation details.

    \item The signed butterfly based methods (\ie SCsc, MFwBT, and SBRW) outperform Deepwalk by 33.0\%, 30.1\%, and 32.2\% on AUC in U.S. House, achieve 28.9\%, 24.3\% and 28.5\% gains on AUC in U.S. Senate, respectively.
    It shows that modeling the balance theory in the signed bipartite network is key for \lsp.
    But in Review and Bonanza datasets, the signed butterfly based methods cannot outperform embedding based methods.
    This can be due to that there are fewer signed butterfly isomorphism in these two datasets.
    Besides, Bonanza is an  extremely unbalanced dataset (\ie \% Positive Links is 0.980), AUC and F1 show a big difference.
    
    \item Our SBGNN model achieve the best results on most metrics.
    Except bonanza, SBGNN-$\textsc{Mean}$ and SBGNN-$\textsc{Gat}$ significantly outperform SCsc.
    In bonanza, SBGNN-$\textsc{Gat}$ significantly gains better results in Binary-F1, Macro-F1, and Micro-F1.
    It demonstrates that SBGNN effectively models signed bipartite networks.
    Besides, we can find that $\textsc{Gat}$ aggregator is better than $\textsc{Mean}$ aggregator, which can be attributed to the attention mechanism~\cite{vaswani2017attention}.    
    
\end{itemize}

\subsection{Parameter Analysis and Ablation Study}

In this subsection, we conduct parameter analysis and ablation study for our SBGNN model.
We choose the U.S. House as our dataset and select 85\% training edges, 5\% validation edges, and 15\% test edges as before. 

\begin{figure}[!t]
\hspace{-0.02\linewidth}
  \begin{center}
       \begin{subfigure}[t]{0.49\linewidth}
    \includegraphics[width=\linewidth]{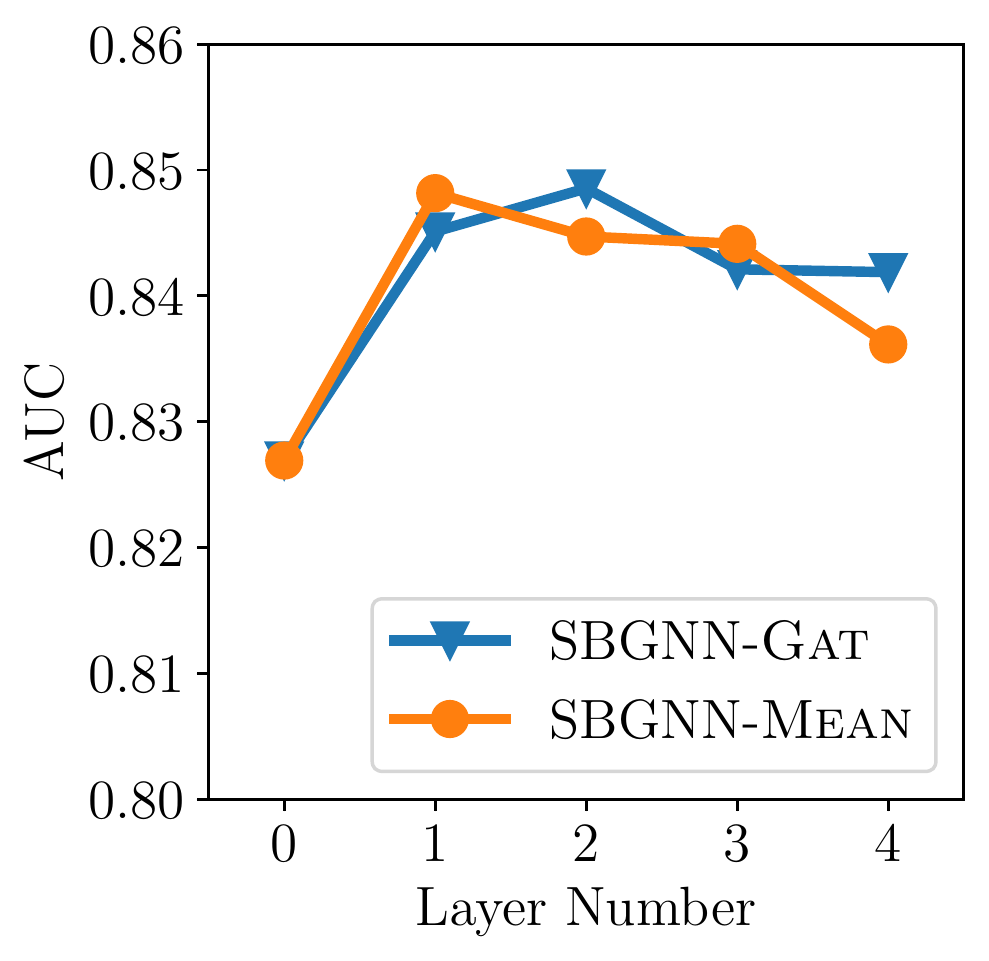}
    \caption{\#Layer $l$}
    \label{fig:layer}
  \end{subfigure}
  \begin{subfigure}[t]{0.49\linewidth}
    \includegraphics[width=\linewidth]{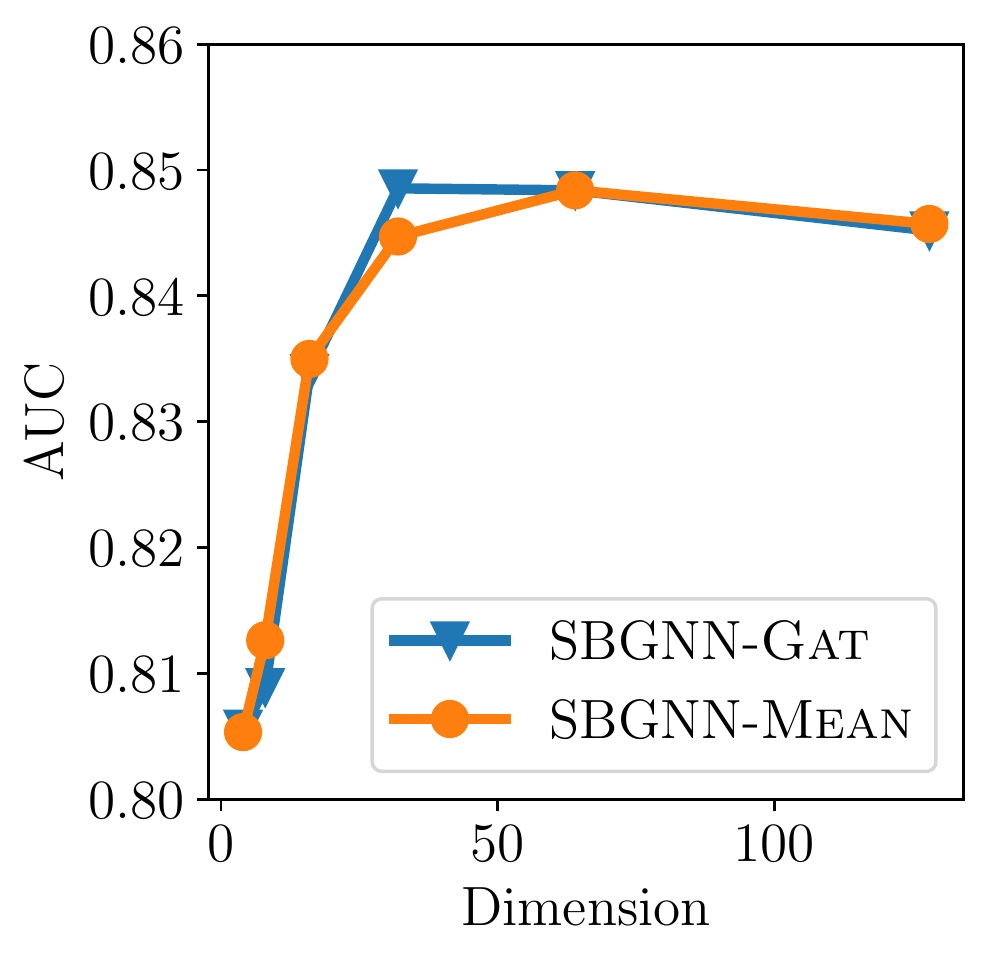}

    \caption{Dimension $d$}
    \label{fig:d}
  \end{subfigure}
  \caption{Parameter analysis on the number of SBGNN Layer $l$ and imension $d$ for SBGNN on the U.S. House dataset.}
  \label{fig:parameter-analysis}
  \vspace{-1em}
  \end{center}
 \end{figure}
 
\subsubsection{Parameter Analysis}
We analyze the number of our SBGNN Layer $l$ and the dimension $d$ of embeddings.
For the number of our SBGNN Layer $l$, we vary $l$ from $\{0, 1, 2, 3, 4\}$. 
Note that, $l=0$ means there is no GNN Layer is used (\ie just lookup embeddings is used), so the results for SBGNN-$\textsc{Mean}$ and SBGNN-$\textsc{Gat}$ are same.
For \figref{fig:layer}, we can find that GNN Layer is more effective than the lookup embedding methods.
For SBGNN-$\textsc{Mean}$, the best $l$ is 1, which AUC is 0.8481. 
But for SBGNN-$\textsc{Gat}$, two SBGNN Layers will get a better results.
 For the dimension of SBGNN, we choose the value $d$ from $\{4, 8, 16 $, $32, 64, 128\}$ to analyze the effects of dimensions.
 From \figref{fig:d}, we can find that as the value increases from 4 to 32, AUC on SBGNN-$\textsc{Gat}$ increases from 0.8056 to 0.8485; AUC on SBGNN-$\textsc{Mean}$ increases from 0.8053 to 0.8447.
 After 32 for SBGNN-$\textsc{Mean}$ and 64 for SBGNN-$\textsc{Gat}$, the AUC value slightly descrease.
 This result can be due to the reason that large dimension $d$ will cause the difficulties of training embeddings.
 
\begin{table}
    \begin{center}
    \caption{Ablation study results for SBGNN model on the U.S. House dataset.}
    \label{tab:ablation_study}
     \scalebox{0.92}{
     
    \begin{tabular}{c|cccc}
    \toprule
    Method & AUC & Binary-F1 & Macro-F1 & Micro-F1 \\
    \midrule
    SBGNN-$\textsc{Gat}$  & 0.8485 & 0.8586 & 0.8477 & 0.8485 \\
    SBGNN-$\textsc{Gat}$ (w/o \textit{Set}$_1$) &  0.8406 & 0.8521 & 0.8400 & 0.8409\\
    SBGNN-$\textsc{Gat}$ (w/o \textit{Set}$_2$) & 0.8440 & 0.8567 & 0.8438 & 0.8448 \\
    SBGNN-$\textsc{Gat}$ (with \textsc{Lr}) & 0.6281 & 0.6195 & 0.6227 & 0.6227\\
    SBGNN-$\textsc{Gat}$ (with \textsc{Mlp}) & 0.8365 & 0.8480 & 0.8358 & 0.8367 \\
    
    \midrule
    SBGNN-$\textsc{Mean}$ & 0.8447 & 0.8519 & 0.8429 & 0.8434\\
    SBGNN-$\textsc{Mean}$ (w/o \textit{Set}$_1$) & 0.8419 & 0.8496 & 0.8402 & 0.8408\\
    SBGNN-$\textsc{Mean}$ (w/o \textit{Set}$_2$) & 0.8296 & 0.8410 & 0.8288 & 0.8297 \\
    SBGNN-$\textsc{Mean}$ (with \textsc{Lr}) & 0.6285 & 0.6387 & 0.6263 & 0.6267 \\
    SBGNN-$\textsc{Mean}$ (with \textsc{Mlp}) & 0.8443 & 0.8531 & 0.8430 & 0.8436\\
    \bottomrule
    \end{tabular}
    \vspace{-4em}
    }
    \end{center}
\end{table}

\subsubsection{Ablation Study}

For the ablation study, we investigate the effect of different aggregation and prediction functions.
Firstly, as we discussed in \secref{sec:experiments-res}, $\textsc{Gat}$ aggregator is better than $\textsc{Mean}$ aggregators.
We further investigate the effect for our message function design.
From \tableref{tab:ablation_study}, we can see that without message from \textit{Set}$_1$ (\ie w/o \textit{Set}$_1$), SBGNN-$\textsc{Gat}$ and SBGNN-$\textsc{Mean}$ descrease 0.9\% and 0.3\%, respectively; without \textit{Set}$_2$ message (\ie w/o \textit{Set}$_2$), SBGNN-$\textsc{Gat}$ and SBGNN-$\textsc{Mean}$ descrease 0.5\% and 1.8\%, respectively.
It demonstrates that both message from \textit{Set}$_1$ and \textit{Set}$_1$ is useful for the SBGNN model.
As we discussed in \secref{sec:loss_function}, the prediction can be achieved by product operation or $\textsc{Mlp}$. 
We replace it with a simple \textsc{Lr} layer or a two-layer $\textsc{Mlp}$.
From \tableref{tab:ablation_study}, we can find that $\textsc{Mlp}$ is much better than simple \textsc{Lr} but not better than product operation.

\section{Conclusions and Future Work}
\label{sec:conclusion}

In this paper, we focus on modeling signed bipartite networks.
We first discuss two different perspectives to model the signed bipartite networks.
Through sign construction, the new perspective can count the signed triangles in the same node type networks.
It obtains consistent results with signed butterfly analysis.
We further use these two perspectives to model peer review and find that after rebuttal, the balance of reviewers’ opinions improved. 
It shows that the rebuttal mechanism makes the reviewer's opinions more consistent.
Under the definition of a new perspective, we propose a new graph neural network model SBGNN to learn the node representation of signed bipartite graphs.
On four real-world datasets, our method has achieved  state-of-the-art results.
Finally, we conducted parameter analysis and ablation study on SBGNN.

In future work, we will explore signed bipartite networks with node features, which can improve the \lsp~\cite{karimi2019multi} and node classification~\cite{tang2016node}.
For example, in the prediction of bill vote, if the node features can be modeled, such as the political standpoint of the congress, it will be more effective in predicting the vote results.
Besides, we will also try to introduce signed bipartite graph neural networks into recommender system~\cite{tang2016recommendations}.

\begin{acks}
This work is funded by the National Natural Science Foundation of China under Grant Nos. 62102402, 91746301 and U1836111. Huawei Shen is also supported by Beijing Academy of Artificial Intelligence (BAAI) under the grant number BAAI2019QN0304 and K.C. Wong Education Foundation.
\end{acks}

\bibliographystyle{ACM-Reference-Format}
\bibliography{refs}

\end{document}